# Sparse incomplete representations: A novel role for olfactory granule cells

Alexei A. Koulakov[1] and Dmitry Rinberg[2]

[1]*Cold Spring Harbor Laboratory, Cold Spring Harbor, NY 11724*
[2]*HHMI Janelia Farm Research Campus, HHMI, Ashburn, VA 20147*

Mitral cells of the olfactory bulb form sparse representations of the odorants and transmit this information to the cortex. The olfactory code carried by the mitral cells is sparser than the inputs that they receive. In this study we analyze the mechanisms and functional significance of sparse olfactory codes. We consider a model of olfactory bulb containing populations of excitatory mitral and inhibitory granule cells. We argue that sparse codes may emerge as a result of self organization in the network leading to the precise balance between mitral cells' excitatory inputs and inhibition provided by the granule cells. We propose a novel role for the olfactory granule cells. We show that these cells can build representations of odorant stimuli that are not fully accurate. Due to the incompleteness in the granule cell representation, the exact excitation-inhibition balance is established only for some mitral cells leading to sparse responses of the mitral cell. Our model suggests a functional significance to the dendrodendritic synapses that mediate interactions between mitral and granule cells. The model accounts for the sparse olfactory code in the steady state and predicts that transient dynamics may be less sparse.

## INTRODUCTION

Mammalian olfactory bulb is the first stage in the processing of information about odorants. The surface of olfactory bulb is covered by several thousands of glomeruli. Each glomerulus receives inputs from a set of sensory neurons expressing the same type of olfactory receptor protein. The inputs into each glomerulus are therefore substantially correlated (Shepherd et al., 2004; Lledo et al., 2005). The glomerulus-based modularity is preserved further by the mitral cells. Most of them receive direct excitatory inputs from a single glomerulus only. Mitral cells are also a major output class of the olfactory bulb that carries information about odorants to olfactory cortex (Figure 1).

In 1950, Adrian reported that the spontaneous activity of the mitral cells in awake rabbit is much higher than that in an anesthetized animal, and in awake animal odor responses vanishe on the background of this spontaneous activity (Adrian, 1950). These observations were confirmed and more thoroughly analyzed in mice in recent study by Rinberg et.al. (Rinberg et al., 2006). The odor responses of mitral cells are sparse and state-dependent (Kay and Laurent, 1999; Rinberg et al., 2006; Fuentes et al., 2008). Only a small fraction of mitral cells responds to the individual odorants despite receiving substantial odorant related input.

How can sparseness emerge mechanistically? One of the mechanisms for the generation of sparse responses involves cells' firing near threshold (Rozell et al., 2008). If the inputs into a cell fall below threshold, the cell does not fire, which leads to the sparse responses. However, this explanation does not apply in the case of mitral cells, because they exhibit high spontaneous firing rates in the awake behaving animals, i.e. specifically in the state when the responses are most sparse.

It is common to explain sparse codes as satisfying the constraint of metabolic energy efficiency (Lennie, 2003; Olshausen and Field, 2004). Although the information content of sparse representations is poor compared to the full codes (Baraniuk, 2007), sparse codes can use less metabolic energy. However, the high spontaneous rates coexisting with sparse representations makes the olfactory code both informationally poor and metabolically costly. The functional significance of sparse olfactory codes is therefore unclear.

These arguments point to neurons other than mitral cells as likely sources of sparseness. Amongst the bulbar neurons, the granule cells (GC) are thought to implement lateral inhibition between mitral cells belonging to different glomeruli through a mechanism based on dendro-dendritic reciprocal synapses (Figure 1) (Shepherd et al., 2004). Such interactions are thought to facilitate discriminations between similar stimuli and mediate competitive interactions between coactive neurons (Arevian et al., 2008). Because GCs are the most abundant cell type of the olfactory bulb it is likely that they are substantially involved in these tasks.

In this study we propose a novel role for the GCs of the olfactory bulb. We show that granule cells can form complimentary to mitral cells representations of olfactory stimuli. Exact balance between the excitation from receptor neurons and the inhibition from the granule cells eliminates odor responses in some mitral cells. However, some mitral cells retain the ability to respond to odors, due to incompleteness of the granule cells representation. The olfactory code on the level of mitral cells becomes sparse. This function is facilitated by the network architecture based on dendrodendritic reciprocal synapses between the mitral and granule cells.

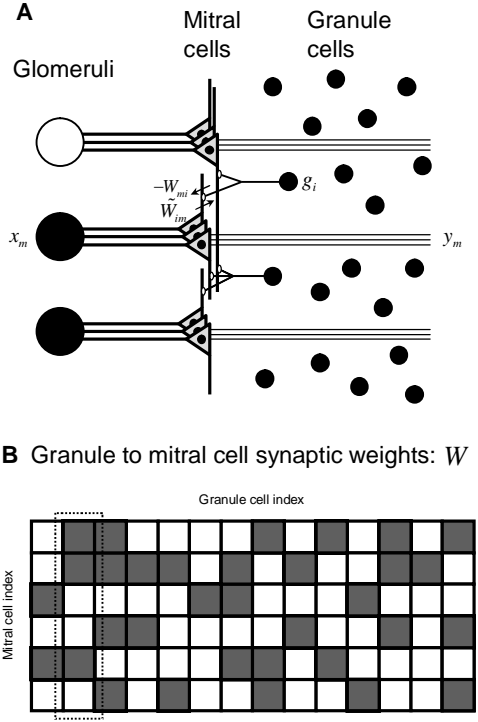

**Figure 1.** Mitral – granule cell network model. **A.** Mitral cells (gray triangles) receive excitatory inputs, $x_m$, from glomeruli, (large circles). The mitral cell output, $y_m$, is sent to the brain. Mitral cells and granule cells (small solid circles) are connected by reciprocal dendrodendritic synapses. Granule cells receive the excitatory input from the mitral cells, which defined by synaptic weight matrix $\tilde{W}_{im}$, and mitral cells are inhibited by granule cells, with weight matrix $W_{mi}$. **B.** Granule to mitral cell synaptic weight $W_{mi}$. Here we argue that GCs represent mitral cell inputs as a superposition of columns in the weight matrix, such as the one marked by the dotted rectangle.

# RESULTS

Our goal is to describe the behavior of two ensembles of neurons: mitral and granule cells (Figure 1A). Mitral cells are projection neurons that receive inputs from olfactory glomeruli and send information about odorants to olfactory cortices for further processing. Mitral cell inputs from the glomeruli are contained in vector $x_m$ (here and below index $m$ represents the mitral cell number running between 1 and $M$, the total number of mitral cells). These inputs represent synaptic inputs from the receptor neurons. The output firing rate of the mitral cells is denoted by $y_m$. The purpose of the present study is to understand the relationship between mitral cell inputs $x_m$ and their outputs $y_m$ in the presence of granule cell inhibition.

GCs are inhibitory interneurons. GC responses are described by the vector $a_i$ with the index $i$ running between 1 and $N$, the number of granule cells. We assume that the granule cells are much more abundant than mitral cells, i.e. $N \gg M$ (Tolley and Bedi, 1994). The mitral cells receive inhibitory inputs from the granule cells determined by the weight matrix $W_{mi}$ (Figure 1B). Each element of matrix $W_{mi}$ describes the strength of synapses between a granule cell number $i$ and a mitral cell number $m$. Because of the relative abundance of granule cells, this matrix is rectangular as shown schematically in Figure 1B. Similarly, the reverse synaptic weights from the mitral to granule cells are described by the matrix $\tilde{W}_{im}$. Two reciprocal weight matrices, $W_{mi}$ and $\tilde{W}_{im}$, although attributed to the same dendrodendritic synapse, are not necessarily the same.

**The system of mitral cells and granule cells connected by reciprocal dendro-dendritic synapses can be described by a Lyapunov function.** The behavior of the network can be simplified if the synaptic weights from mitral cells to granule cells $\tilde{W}_{im}$ are proportional to the reverse synaptic weights from granule cells $W_{mi}$:

$$W_{mi} = \varepsilon \tilde{W}_{im}. \qquad (1)$$

This assumption is not unrealistic because both weights are determined by active zones within the same dendrodendritic synapse (Shepherd et al., 2007), and can, in principle, be regulated to be proportional. In the Methods section we show that under this condition the dynamics of the network is described by minimization of the cost function that is called Lyapunov function (Hertz et al., 1991). Lyapunov function is a standard construct in the neural network theory that has been extensively used to study the properties of complex networks. The steady states of the neural activities can be obtained as minima of this function. Thus, by finding minima of the Lyapunov function one can understand the steady-state responses of mitral and GCs to odorants. The Lyapunov function for mitral - GC network has the following form:

$$\mathcal{L} = \frac{1}{2\varepsilon} \sum_{m=1}^{M} \left( x_m - \sum_i W_{mi} a_i \right)^2 + \sum_{i=1}^{N} C(a_i). \quad (2)$$

The first term in the function contains the sum of squared differences between the excitatory inputs to the mitral cells from receptor neurons $x_m$ and the inhibitory inputs from the granule cells. The inhibitory inputs are proportional to the activity of granules cells $a_i$ weighted by synaptic matrix $W_{mi}$. The first term therefore reflects the balance between excitation and inhibition on the inputs to mitral cells. Minimization of this term due to the system's dynamics leads to the establishment of an exact balance between excitation and inhibition. The second term of the Lyapunov function represents a cost imposed on the granule cell firing. It reaches minima when $a_i = 0$, which correspond to minimization of number of active GCs.

**Granule cells represent mitral cell inputs as a superposition of dendrodendritic weights.** We will first assume that the second term in the Lyapunov function is negligible. This condition corresponds to the case when the granule cells are easy to activate and, therefore, the cost associated with their firing is small. Minimizing the first term of the Lyapunov function without the cost leads to the precise balance between excitation $x_m$ and inhibition $\sum_i W_{mi} a_i$ [see equation (2)]:

$$x_m = \sum_i W_{mi} a_i. \quad (3)$$

According to this equation the dendrodendritic weights can be viewed as vectors $\vec{W}_i$ defined in the space of mitral cell inputs (gray arrows in Figure 2A). The firing rates of granule cells $a_i$ can be viewed as unknown coefficients. The problem posed by equation (3) can be restated as follows: Represent the vector of mitral cell excitatory inputs $\vec{x}$ as a superposition of vectors $\vec{W}_i$ by finding the coefficients $a_i$ (Figure 2). We conclude that the network dynamics in olfactory bulb leads to the representation of the bulbar inputs by the granule cells using the dendrodendritic synaptic weights as basis vectors. In this model the granule cells play the role of coding cells that carry representations of mitral cell inputs.

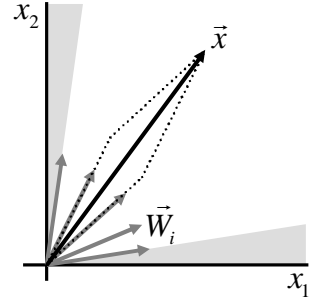

**Figure 2.** The olfactory system with two mitral and five GCs. The inputs into mitral cells from the receptor neurons are represented by the 2D vector $\vec{x}$. The weights from the GCs are shown by five 2D vectors $\vec{W}_i$. Equation (3) leads to the representation of the input vector in terms of the weight vectors, as shown by the dashed lines, with the coefficients given by the activities of the GCs. In this example, the second and the third GCs are active, although other solutions are also possible.

The number of the basis vectors $\vec{W}_i$ is equal to the number of granule cells, $N$. The dimensionality of the input vector $\vec{x}$ is equal to a number of the mitral cells, $M$, which is much smaller than $N$. Therefore, the problem of representation of mitral cell inputs as a superposition of basis vectors may have several solutions. Thus, in Figure 2 five granule cells (gray arrows) have to represent inputs into two mitral cells ($x_1$ and $x_2$). Clearly there are many ways how a 2D vector can be decomposed into five components.

This picture suggests the following interpretation of the role of the bulbar GC. The strength of dendrodendritic synapses represents the receptive fields of individual GCs, i.e. specific features in the glomerular inputs that these cells respond to. The bulbar network attempts to decompose its inputs into a superposition of these individual features. Because of the abundance of the GCs, no unique representation of the inputs in terms of GC receptive fields can be found. The olfactory code as defined by the GCs is therefore underdetermined. As suggested by the studies of underdetermined (overcomplete) systems the unique solution can be found if one includes the cost imposed on the firing rate vector into consideration. This is the second term in equation (2) that we so far neglected.

**Mitral cell firing rates result from the balance between excitation and inhibition.** In our model the response of a mitral cell is defined as the deviation of the firing rate of a mitral cell, $\Delta y_m$, from spontaneous rate, $\bar{y}_m$:

$\Delta y_m = y_m - \bar{y}_m$. This response results from the excitatory inputs from receptor neurons $x_m$ and inhibitory inputs from the granule cells

$$\Delta y_m = x_m - \sum_{i=1}^{N} W_{mi} a_i . \quad (4)$$

Therefore the response $\Delta y_m$ of the mitral cell is a result of the balance between excitatory and inhibitory inputs. If such a balance is established precisely, i.e. if the granule cells represent exactly the bulbar inputs according to equation (3), no mitral cell is expected to respond to odorants. On one hand, this observation implies that the responses of mitral cells are ultimately sparse - they completely disregard their excitatory inputs. This model therefore contains the feature of the olfactory code observed experimentally, i.e., sparseness. Sparseness in this model emerges as a result of self-organizing balance between excitation and inhibition. On the other hand, it is hard to imagine a utility of this type of code, because the mitral cells do not respond to odorants at all. Next we identify the features of the model that could allow mitral cells to respond to odorants.

**Sparse mitral cell responses result from incomplete granule cell representations.** How can responses of mitral cells emerge in our model? If GCs represent the excitatory inputs of the mitral cells exactly, mitral cells are unresponsive. Therefore GCs have to fail to represent glomerular inputs exactly for the mitral cells to respond to odorants. The GC code therefore has to be *incomplete* leading to an inaccurate odorant representation by the GCs.

Here we present the simplest example of such an incomplete GC code. In Figure 3 we illustrate the synaptic currents emerging in the simplistic olfactory system with six mitral cells and only one GC. Vector $\vec{x}$ represents the excitatory inputs received by all six mitral cells. In Figure 3A the odorant supplies inputs into three mitral cells out of six. We then assume that the granule-mitral cell weights $\hat{W}$ match exactly this input pattern. As a result, when the GC is activated, it will return an inhibitory input $\vec{\hat{x}}$ into to the mitral cells that exactly matches the pattern of excitation. The responses on the mitral cells to the odorant will be very weak as a result of such a compensation (some small residual responses are needed to drive the GC).

Figures 3B and C show examples of two odorants that activate similar subsets of mitral cells. The pattern of inhibition $\vec{\hat{x}}$ however is exactly the same as in Figure 3A, because with the single GC present, the inhibitory currents are constrained to a single pattern. As a result of subtraction between receptor neuron excitatory and GC inhibitory inputs into the mitral cells, only a single mitral cell responds to each of these odorants. Thus, the responses of mitral cells become sparse. Some mitral cells relay their inputs, while others are almost completely inhibited by the GCs. That some mitral cells respond is a consequence of the incompleteness of the GC code.

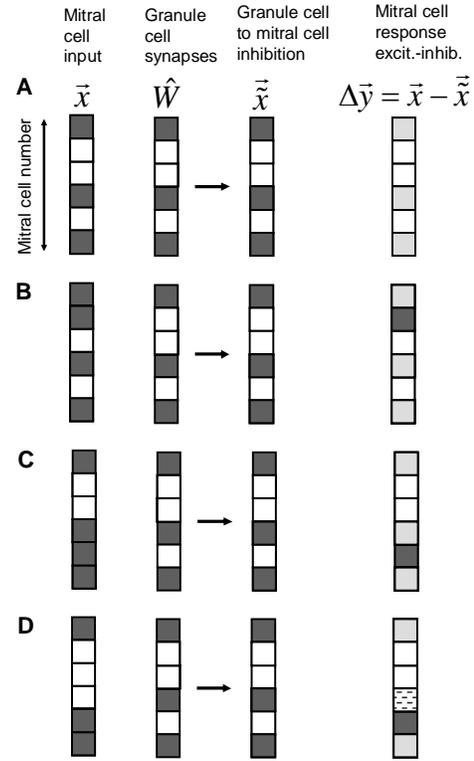

**Figure 3.** Simple olfactory system that contains six mitral and one granule cell. Gray areas indicate strong activity. Light gray areas on the right indicate weak responses due to possible small correction induced by the cost function. (A-D) responses to four different odorants represented by the combinatorial inputs shown on the left.

The result of sparsening of the mitral cell responses is the reduction of redundancy of the odorant representation. Thus, although mitral cells may receive highly redundant inputs (Figure 3B and C, left column), only few may respond (right column). The sparseness in the mitral cell responses may be important for reducing overlaps between odorant representations. In the example in Figures 3B and C, responses of mitral cells to two odorants share no overlap. Figure 3D shows the responses to an odorant that are inhibitory (below the baseline activity). Therefore, the incompleteness in the GC odorant representation may emerge from the restriction on the diversity in the GC weights.

**Granule cell nonlinear input-output relationship may be responsible for sparse granule cell coding.** The example presented above contained only one GC. Therefore GC code was incomplete leading to the mitral cell firing. If the number of GCs was equal or larger than the number of mitral cells, there are enough vectors in the basis to represent any glomerular input. Because the olfactory bulb contains many more GC than mitral cells one should seek some other source for the incompleteness of GC representation. We propose that a source of incompleteness is in the constraint that GC firing rates cannot be negative.

To satisfy non-negativity condition for the GCs firing rate, the cost function $C(a_i)$ in equation (2) should be infinite for negative $a_i$. In general, cost $C(a_i)$ is defined by the non-linearity in the cells' input-output relationship. The simplest example is shown at Figure 4.

For the positive firing rates the cost function increases proportionally to the GC threshold for firing $\theta$, making it more costly to fire for the cells with a large threshold. This feature selects which GC will be active in the stationary state with the higher threshold cells less likely to fire. The exact relationship between the input-output relationship and the cost function is given by equation (11) of the Methods.

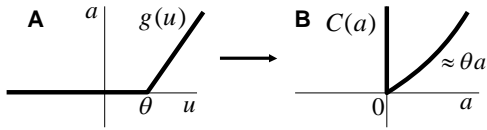

**Figure 4.** Granule cell input-output relationships (A) and the corresponding cost-functions $C(a)$ (B). $u$ is the granule cell's input current. $a$ is the resulting firing rate. The cost function prohibits negative firing rates, which reflects the feature of the input-output relationship. For positive firing rates the cost-function is approximately linear.

Because GC firing rates cannot be negative, some odorants cannot be exactly represented by the GCs. Indeed, consider the gray regions in Figure 2A. The input vector $\vec{x}$ in this area cannot be represented as a superposition of GC weights (gray arrows) with positive coefficients. To exactly represent a vector in this area, at least one GC has to have a negative firing rate, which is not allowed. Therefore, the inputs within gray area will lead to an *incomplete* representation by the granule cells. The exact balance between inhibition and excitation on the input of some mitral cells will be destroyed and, according to equation (4), some mitral cells will respond to these inputs.

A more high-dimensional example is shown in Figure 5. From this example it is clear that substantial responses of mitral cells emerge when the input from receptor neurons falls outside the conical domain spanned by the GC weight vectors (gray vectors in Figure 5B). In this case the inhibitory currents from the granule cells to the mitral cells are given by the nearest vector $\vec{\tilde{x}}$ on conical domain that can be represented as a superposition of GC weights with positive coefficients. The response of the mitral cells to the odorant it the result of the difference between excitatory receptor neuron inputs $\vec{x}$ and inhibitory inputs from the granule cells $\vec{\tilde{x}}$. Thus, substantial mitral cell response is possible if $\vec{x} \neq \vec{\tilde{x}}$, i.e. there is no precise balance of inhibition and excitation on the input to mitral cells. In this case the representation of the receptor neuron inputs by the GC is incomplete.

The responses of GC are sparse for the incomplete representations. Thus, in Figure 5B, for the 3D space characterizing inputs into three mitral cells only two GCs out of eight (red arrows) are active. This is because GCs approximate the vector lying on the surface of the region spanned by GC weights that is 2D. In multidimensional space, when $M$ mitral cells are present, the active GCs weight vectors lie on the surface of the domain spanned by all GCs, so the number of active GCs is smaller or equal $M-1$.

Because the number of GCs is much larger that the number of mitral cells $N \gg M$, one expects that responses of GCs are very sparse. This reasoning is confirmed more rigorously in the Theorem 1 of Methods.

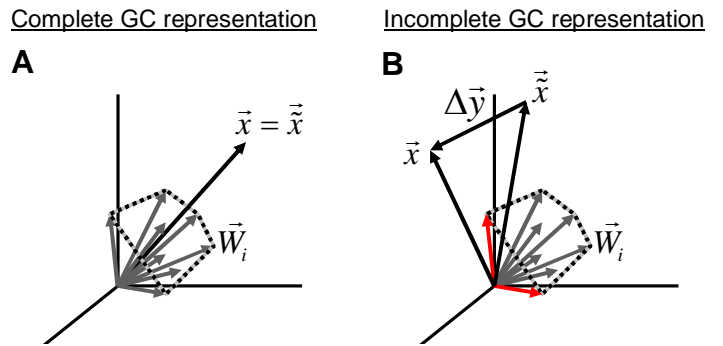

**Figure 5**. The olfactory bulb with three mitral cells (3D input space) and eight GCs. The inputs into mitral cells from receptor neurons are represented by the black arrows. The synaptic weights from eight GCs to the mitral cells are shown by the gray arrows. (A) If inputs lie within the cone restricted by the weight vectors (dashed) the input vector can be represented as a superposition of weight vectors exactly. The GC representation of mitral cell inputs is exact and the mitral cells are expected to respond weakly due to exact balance between inhibition and excitation. (B) If the input vector is *outside* the cone of GC weights (dashed) the GCs cannot represent the inputs precisely. This is because the GC firing rates (coefficients of expansion) cannot be negative. The best approximation of the inputs is shown by vector $\vec{\tilde{x}}$. Vector $\vec{\tilde{x}}$ is the nearest vector on the cone to the input vector $\vec{x}$. This vector is formed by two GCs (red). Because $\vec{x}$ differs from GC representation $\vec{\tilde{x}}$, the GC code is *incomplete*. Incompleteness of representation leads to substantial mitral cell firing $\Delta \vec{y} = \vec{x} - \vec{\tilde{x}}$.

**Incomplete granule cell representations are typical for random network weights.** It may appear that incomplete representations are limited to the edges of the parameter space, such as the gray area in Figure 2A. The situation shown in Figure 5A may seem to be more typical than that in Figure 5B. Here we argue that this intuition applies only to input spaces of small dimensionality such as shown in Figures 2 and 5. For a multidimensional input space, which corresponds to large number of mitral cells, random inputs and network weights create incomplete representations. This implies that it becomes almost impossible to expand a random input vector with positive components into the basis containing vectors with positive components using only non-negative coefficients without the loss of precision. In the Methods section we show that the number of coactive GC for random binary inputs with $M$ mitral cells is $\sim \sqrt{M}$. Because a precise representation of the $M$-dimensional random input requires $M$ vectors, this result implies that the representation of odorants by the GCs is typically imprecise. The GC code is therefore incomplete. We also show that for sparse GC-to-mitral cell connectivity, when only $K \ll M$ weights are non-zero, the number of coactive GC is somewhat larger $\sim M/\sqrt{K}$. This number is still substantially smaller that what is required for complete GC code. We conclude that GC cannot represent mitral cell inputs precisely for the case of random connectivity, which implies ubiquity of incomplete representations.

**The state dependence of the granule cell code.** Above we have discussed two reasons for the incompleteness of GC representation. First, the incompleteness may result from the restricted diversity of the GC weights (Figure 3). Second, we suggested that the non-negativity of the GC firing rates may lead to an inaccurate representation of odorants implying incompleteness. Here we discuss the third reason why GC may yield an inaccurate code. We suggest that the presence of finite (non-zero) threshold for GC firing (Figure 4A) can lead to an inaccurate representation. This phenomenon allows us to explain the transition between the awake and anesthetized responses.

To demonstrate our point we will use simplified model of bulbar network containing only one granule cell (Figure 3) This network has an advantage that an exact solution can be found and understood even when a finite threshold for firing is present in the activation of the GCs. Indeed, consider input configuration shown in Figure 3B. Assume for simplicity that all of the non-zero weights and mitral cell inputs have unit strengths. Then, the Lyapunov function for the activity of the single GC $a$ is

$$\mathcal{L}(a) = \frac{K}{2}(1-a)^2 + \theta a. \quad (5)$$

Here, of course, we have to assume that $a \geq 0$. $K$ is the number of non-zero weights for the GC ($K = 3$ in Figure 3). By minimizing the Lyapunov function we obtain $a = 1 - \theta/K$, for $\theta \leq K$, and zero otherwise. The activity of the mitral cell number two (Figure 3B and 6) cannot be affected by the GC, because the latter makes no synapses onto this cell. The activities of mitral cells 1, 4, and 6 are given by

$$\Delta y_{1,4,6} = 1 - a = \frac{\theta}{K} \quad (6)$$

for $\theta \leq K$. This formula means that the mitral cells will increase their firing rate slightly to activate the GC. The amount of increase is equal to the threshold for activation of the GC divided by the number of mitral cells contributing to the input current, i.e. $K$. The activity of the GC is also assumed to raise fast above the threshold so that it suppresses all firing increases on its input above this value [equation (6)]. The responses of mitral cells as functions of the threshold $\theta$ are shown in Figure 6.

Figure 6 shows that for large thresholds $\theta$ all mitral cells that receive receptor inputs will respond to the odorant. In this regime the GC is silent due to the high threshold and, therefore, the mitral cells relay the receptor neuron inputs directly. The mitral cell code is not sparse in this case. We argue that the case of large thresholds corresponds to the network in the anesthetized animal. When the threshold for GC firing is small, the mitral cell firing becomes sparse (Figure 6). This is the regime considered throughout this paper, which, we argue, corresponds to the awake animal. According to this model therefore the transition from awake to anesthetized state is accomplished by an increase in the thresholds of GC firing, which could be mediated by the centrifugal projections.

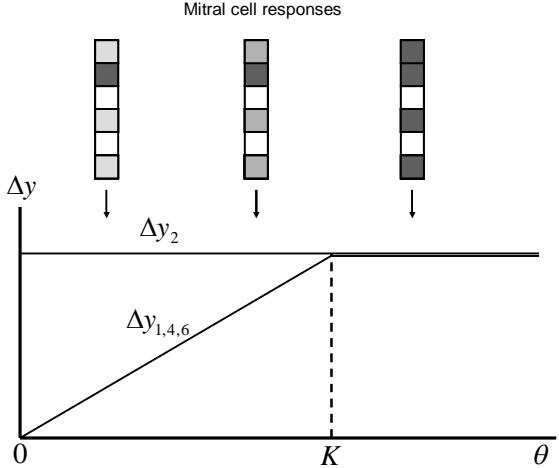

**Figure 6.** Sparsification of responses as a function of parameter $\theta$ (threshold for GC firing). Example shown in Figure 3B is used. We argue that for large thresholds the response of mitral cells is less sparse than for small thresholds. Because the threshold for GC firing can be regulated by the centrifugal inputs from cortex, increase in firing threshold may mediate the transition between awake and anesthetized states.

## DISCUSSION

**Sparse overcomplete codes**. At the core of these representations is an assumption that a sensory input can be decomposed into a superposition of primitives that are called dictionary elements (Field, 1994; Olshausen and Field, 2004; Rozell et al., 2008). The decomposition is sought in the form of a set of coefficients with which different dictionary elements contribute to the input. These coefficients are thought of as responses of neurons in a high level sensory area that indicate whether given feature is present in the stimulus or not. Because the number of primitives available is usually large, several decompositions are consistent with the inputs. If the set of primitives is overcomplete, the decomposition problem generally has no unique solution. To make sensory representation unambiguous some form of parsimony principle is added to the model in the form of cost function on the coefficients/responses. The solution that yields a minimum of the cost function is assumed to be chosen by the nervous system. The decomposition is found to be dependent on the cost function. The general form of a cost function is a sum of firing rates in power $\alpha$: $L_\alpha = \sum_i |a_i|^\alpha$. For $L_2$, a simple sum of squared coefficients, it can be shown that generally all neurons will respond to any stimulus, and, therefoore, the code is not sparse. For $L_1$ norm when the cost is equal to the sum of absolute values of the responses, the solution is found to be sparse. This means that only a few neurons will respond to a stimulus. Thus, $L_1$ norm is called sparse and the corresponding neural representation is called sparse overcomplete.

It was found that a network of inhibitory neurons can implement sparse overcomplete codes (Rozell et al., 2008). To show this, the dynamics of the system was represented as minimization of a cost function called a Lyapunov function. The basic idea of this method goes back to John Hopfield, who was the first to evaluate the Lyapunov function for a recurrent network (Hopfield, 1982, 1984). Hopfield network has attractor states that allow to implement associative memory (Hopfield, 1982; Hertz et al., 1991). In contrast to Hopfield networks, in purely inhibitory networks, the recurrent weights entered the Lyapunov function with a minus sign, which abolished the attractor memory states and made the network purely sensory (Rozell et al., 2008). Minimization of Lyapunov function in realistic recurrent networks with inhibition was suggested to implement the parsimony constraint mentioned above.

To implement sparse overcomplete representations with realistic networks of neurons at least two requirements on the network connectivity have to be met (Rozell et al., 2008). First, the feedforward weights between the input layer of the network and the inhibitory neurons have to contain the dictionary elements (Figure 7A). Thus, the inhibitory neurons that are responsible for a particular dictionary element will be driven strongly when it is present in the input, due to high overlap between the stimulus and the feedforward weight. Second, the recurrent inhibitory weight between any pair of neurons has to be proportional to the overlap between their dictionary elements (Figure 7A). Because inhibition between neurons implements competition, this feature implies that similarly tuned neurons compete most strongly. Therefore, the two types of network weights, feedforward and recurrent, have to be in close match, one of them constructed as the overlap of the other. It is not clear how this match is established in a biological system, because these are different sets of synapses. Here we argue that the olfactory bulb network architecture that is based on dendrodendritic synapses can ensure that the feedforward and recurrent connectivity are in close match.

## Sparse overcomplete representation networks

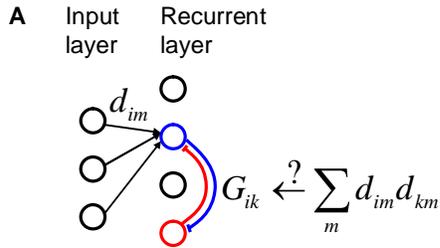

## Olfactory bulb

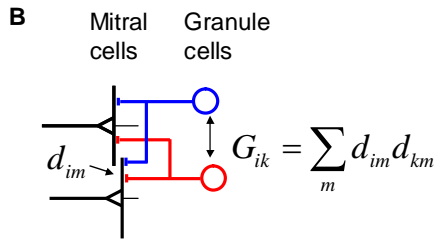

**Figure 7**. A possible functional significance of dendro-dendritic synapses. (A) Networks implementing sparse overcomplete representations. These networks decompose the input pattern into a parsimonious superposition of dictionary elements. Response of each cell in the recurrent layer indicates whether given dictionary element is present in the input. The feedforward excitatory weights contain the dictionary elements. The recurrent inhibitory weights are proportional to the overlap between dictionary elements. More similar input patterns compete stronger. It is not clear however how this condition on the recurrent weights is implemented biologically. (B) In the networks with dendrodendritic synapses the connectivity between granule cells is inhibitory. Because every two granule cells are two synapses away from each other, the strength of pair wise inhibition contains overlap between dendrodendritic weights thus automatically satisfying the constraint on the recurrent inhibitory weights shown in (A).

**Dendrodendritic synapses facilitate the representation of odorants by the granule cells.** GCs of the olfactory bulb receive excitatory inputs from the mitral cells through the dendrodendritic synapses (Shepherd et al., 2004). These synapses encode patterns that can drive strongly individual GCs. At the same time the effective connectivity between GCs is inhibitory (granule to mitral and mitral to granule synapses are inhibitory and excitatory respectively). Here we argue that the strength of mutual inhibition between GCs may be proportional to the overlap between their input weights. The reason for this is that individual GCs are two synapses away from each other. To calculate their mutual inhibition one has to sum over all intermediate synapses, which leads to evaluation of a convolution or overlap (Figure 7B). Therefore, GCs inhibit each other stronger if their inputs from mitral cells are more similar. Thus, both requirements necessary to implement sparse overcomplete representations stated previously are met. This implies that the function of GCs may be to detect specific patterns of activity in the inputs that mitral cells receive. The GCs then build the parsimonious representation of mitral cell inputs. The parsimony is ensured by the mutual inhibition between GCs, with more similar GCs inhibiting each other stronger. The latter condition is facilitated by the network architecture based on dendrodendritic synapses. This observation provides a potential explanation for the existence of these synapses that so far have been discovered in the olfactory bulb only (Shepherd et al., 2004).

**Mitral cells respond to odorants because granule cell code is incomplete.** Two problems emerge if we assume that granule cells implement sparse overcomplete codes. First, granule cells are interneurons and, as such, cannot directly transmit their representation to the upstream network. The significance of these representations becomes unclear. Second, we find that if granule cells establish an absolutely accurate representation of their inputs, the mitral cells will respond to odorants very weakly. This is because granule cells can eliminate these responses from the mitral cells' firing. These considerations suggested to us that the representations by the granule cells are incomplete. This implies that granule cells cannot find accurate representations of their inputs, for example, because this would require for the firing rates to become negative. If granule cells' codes are incomplete the mitral cells transmit only the unfinished portion of the representation to the upstream olfactory networks. As a consequence the representations of odorant by the mitral cells becomes sparse. The redundancies in the mitral cell codes are reduced and the overlaps in representations of similar odorants are erased. As a result the patterns of mitral responses to similar odorants become more distinguishable from each other.

**The Lyapunov function.** Lyapunov functions are standard tools in the neural network theory (Hertz et al., 1991). (Seung et al., 1998) have shown that the network that contains two populations of neurons, inhibitory and excitatory, can be described by the Lyapunov function. This model can be mapped onto the system of mitral-GCs. Here we have eliminated the dynamics of mitral cells from consideration and reduced the description to GCs only. By doing so we have shown that GC could be used to extract features from mitral cell inputs. (Lee and Seung, 1997) have proposed a network mechanism that implements conic encoders, i.e. the ones that represent inputs under non-negativity constraint. Although the non-linear network mechanism is somewhat different from the one used here, the set of encoding/error neurons in Lee and Seung can be viewed as analogs of granule/mitral cells, respectively, in our model. In our study we propose a mapping of the conic networks onto the olfactory bulb network and study the implications of this mapping for the olfactory code.

**Feedback from higher brain areas and state dependence of the olfactory code.** The sparseness of the mitral cell responses depends on the nonlinearity of the GCs, and specifically on the GC activation threshold $\theta$. In this study we assumed that GCs have similar activation threshold that is small enough for them to be easily activated by small levels of activity in the mitral cells. If the thresholds for activation of individual GCs are different, one can envision a mechanism how olfactory code carried by both mitral and granule cells can be controlled to adapt to a particular task. Thus, if the threshold for activation is raised for a subset of GCs, these cells will no longer be active and, therefore, their activity will not be extracted from the firing of mitral cells. If, for example, the threshold for all of the GCs is increased thus making them unresponsive, the olfactory code carried by the mitral cells replicates their inputs from receptor neurons. Conversely, if the activation threshold is lowered for a subset of GCs, these cells will efficiently extract their activity from the mitral cells' responses. Thus, a particular redundancy amongst similar odorants can be excluded in a task-dependent manner. Therefore, the thresholds for GC activation may regulate both an overall sparseness of mitral cell responses and the fine structure of the bulbar olfactory code.

GC threshold values depend on cellular properties, but also can be effectively modulated by additional input into these cells. GCs in the mammalian olfactory bulb are recipients of the signals from the efferent projections from the cortex and other brain areas (Davis and Macrides, 1981; Luskin and Price, 1983). Additional signal to GCs can change the effective threshold values. If a GC receives excitatory inputs from the cortex, the mitral cell signal is closer to the threshold value and the granule cell is easier to excite by the odorant-related inputs. These features provide an opportunity to control sparseness and information content of the mitral cell signal by regulating the input to the GCs from the cortex.

Implicit experimental evidence for the presence of such regulation already exists. In the anesthetized state, when the efferent signal coming to the bulb from the cortex is minimal, the mitral cells respond strongly to the odor stimulation. In the awake state, when the cortex is active, the mitral cell code becomes much sparser (Adrian, 1950; Rinberg et al., 2006). Cutting lateral olfactory tract and eliminating feedback from the brain in awake rabbit made mitral cells respond in the similar manner as in an anesthetized rabbit (Moulton, 1963). Therefore centrifugal projection may indeed regulate sparseness of the olfactory code.

Some evidence also points towards the possibility of finer network tuning by specific activation or deactivation of subsets of GCs to enhance extraction of relevant information. First, (Doucette and Restrepo, 2008) demonstrated that the mitral cell responses to odorants change as animals learn the task. Second, Fuentes et.al. show that the number of mitral cell responses depends on the task (Fuentes et al., 2008). When a rat is involved in odor discrimination task the average number of mitral cell excitatory responses is less then in the rat passively smelling an odor. The assumption is that when an animal discriminates odorants, it may be advantageous to suppress redundant mitral cell responses and enhance those, which carry the most behaviorally relevant information. Our model proposes the neuronal mechanism for this phenomenon.

**Transient regime.** In this study we explicitly considered the steady state regime and used Lyapunov function approach to find the response of the mitral and GCs to the stimulus in the stationary state. Obviously the system's response to odorants is dynamic. What happens when the stimulus appears, when an animal first sniffs an odorant? The thorough analysis of this problem is beyond the scope of this paper, however, we would like to present some obvious points, which lead to specific experimental predictions.

If the signal first appears on the input to mitral cells, it causes elevation of the mitral cell firing rates, which in turn causes activation of GCs and suppression of mitral cells by feedback inhibition. The initial responses of the mitral cells are therefore represented by quick bursts of activity followed by decay to the steady state described in this study. The olfactory bulb may transmit information about the stimulus by these quick bursts of activity. This observation leads to several questions. First, what information about the stimulus is sent to the cortex by transient bursts and of activity and the consequent steady state? Second, what are the experimental conditions for observations of such bursts?

While it is hard to say convincingly anything about the role of different modes of information transmission, we may speculate about the second question. The mitral cell bursts of activity presumably are short and stand on top of the high spontaneous activity. In order to observe them reliably one needs to synchronize spike trains with stimulus delivery. In mammals, stimulus delivery is controlled by sniffing. In the previously studies (Kay and Laurent, 1999; Rinberg et al., 2006; Doucette and Restrepo, 2008; Fuentes et al., 2008), the authors synchronized their recordings with stimulus onset, but not with sniffing/breathing. This approach may lead to smearing of the short bursts, and emphasize the steady state responses of the network. In such regime the odor responses should be sparse as predicted by the model. New evidence by (Shusterman et al., 2009) and (Cury and Uchida, 2009) suggests the presence of fast bursts of mitral cell activity synchronized with sniffing.

**Conclusions.** In this study we addressed theoretically the experimental evidence of sparse odor codes carried by the mitral cells in awake behaving rodents. We proposed a novel role for the granule cells of the olfactory bulb in which they collectively build an incomplete representation of the odorants in the inhibitory currents that they return to the mitral cells. Because the representation formed by granule cells is incomplete, mitral cells can carry information to the cortex in the form of sparse codes. This function is facilitated by the network architecture that includes bidirectional dendrodendritic mitral-to-granule cell synapses. We suggest that the synaptic strengths in the individual synapses are correlated between the two directions.

**Acknowledgments.** The authors are grateful to Dmitry Chklovskii, Barak Pearlmutter, and Sebastian Seung for helpful discussions.

## METHODS

**Derivation of the Lyapunov function.**
Our model is based on the following equations describing the responses of mitral and granule cells respectively:

$$\Delta y_m = x_m - \sum_{i=1}^{N} W_{mi} a_i \qquad (7)$$

$$\dot{u}_m = -u_m + \sum_{m=1}^{M} \tilde{W}_{im} \Delta y_m \qquad (8)$$

$$a_i = g(u_i), \qquad (9)$$

where $u_i$ is the membrane voltage of the granule cell and $\Delta y_m = y_m - \bar{y}_m$. We use the following abbreviation for the temporal derivatives $du/dt = \dot{u}$. We assume therefore that the dynamics of mitral cell responses is fast enough to reflect the instantaneous values of inputs. If one defines a function of granule cell firing rates

$$\mathcal{L}(\vec{a}) = \frac{1}{2\varepsilon} \sum_{m=1}^{M} \left( x_m - \sum_i W_{mi} a_i \right)^2 + \sum_{i=1}^{N} C(a_i), (10)$$

where the cost function is defined as

$$C(a) = \int_0^a g^{-1}(a') da', \qquad (11)$$

one can show that the behavior or the system can be viewed as a form of gradient descent, i.e. $\dot{u}_i = -\partial \mathcal{L}/\partial a_i$. The time derivative of the Lyapunov function $\mathcal{L}$ is therefore always negative

$$\dot{\mathcal{L}} = \sum_i \frac{\partial \mathcal{L}}{\partial a_i} \dot{a}_i = \sum_i (-\dot{u}_i) a_i = \\ = -\sum_i \left[ g^{-1}(a_i) \right]' \left[ \dot{a}_i \right]^2 \leq 0 \qquad (12)$$

The last inequality holds if $g(u)$ is a monotonically increasing function. Therefore the Lyapunov function always decreases if the system behaves according to equations (7) through (9). Since the function is limited from below, the stable states of the system are described by the minima of the Lyapunov function.

The cost function for the threshold-linear neuronal activation function $g(u) = [u - \theta]_+$ shown in Figure 4A can be approximated by a linear function when the firing rate of the granule cell are not too large

$$C(a) \approx \theta a, \quad a \geq 0 \qquad (13)$$

For negative values of firing rates $a$ the cost function is infinitely big, reflecting the fact that negative firing rates are not available.

**The importance of condition (1)**
For our model to have a Lyapunov function a more general condition than (1) may hold. Indeed, the sufficient condition for the Lyapunov function to exist is that the network weight matrix $G_{ik} = \sum_i \tilde{W}_{im} W_{mk}$ is symmetric (Hertz et al., 1991). This will be true if, for example, $W_{mi} = \tilde{W}_{in} E_{nm}$ where $E_{nm}$ is an arbitrary symmetric $M$ by $M$ matrix. Thus condition (1) is sufficient but not necessary for the network to have a Lyapunov function. We argue however that this condition is necessary for the system to be described by the Lyapunov function in the form of equation (2).

**The number of coactive granule cells**
Here we will distinguish two types of Lyapunov functions. First, we will consider the homogeneous case, when the Lyapunov function has the form

$$\mathcal{L}_0(\vec{a}) = \frac{1}{2\varepsilon} \sum_{m=1}^{M} \left( x_m - \sum_i W_{mi} a_i \right)^2. \qquad (14)$$

In the homogeneous case we disregard by the contribution of the cost function other than constraining the firing rates to be non-negative. This condition corresponds to vanishingly small threshold for the activation of the granule cells. The non-homogeneous Lyapunov function is

$$\mathcal{L}(\vec{a}) = \mathcal{L}_0(\vec{a}) + \sum_i \theta_i a_i \qquad (15)$$

with the same constraint $a_i \geq 0$. We will prove here two theorems that limit from above the number of coactive granule cells (i.e. the ones for which $a_i \neq 0$). A quantity that will be important to us is the error in approximation $\Delta x_m \equiv x_m - \sum_i W_{mi} a_i$ that is also, according to equation (4), the response of mitral cells.

*Theorem 1: Sparseness of the homogeneous solution*

Assume that the $M$-dimensional receptive fields of the granule cells $\vec{W}_i$ are the vectors of general position, i.e. every subset of $M$ vectors $\vec{W}_i$ are linearly independent. Then, in the minimum of the homogeneous Lyapunov function (14), either $\Delta x_m = 0$ for all $m$ (i.e. mitral cells do not respond, granule cell representation is complete) or fewer that $M$ granule cells are active. In the former case ($\Delta x_m = 0$) all of the granule cells may be active.
*Proof:* assume that $M$ or more GCs are simultaneously active. Let us vary slightly the activity of only one GC: $\Delta a_k = \varepsilon$. The corresponding variation in the Lyapunov function is $\Delta \mathcal{L}_0 = \left(\Delta \vec{x} \cdot \vec{W}_k\right) \varepsilon + O(\varepsilon^2)$. Because we are considering the minimum of the Lyapunov function, all of the scalar products $\left(\Delta \vec{x} \cdot \vec{W}_k\right)$ have to be zero, which is possible only if $\Delta \vec{x} = 0$ or the number of vectors $\vec{W}_k$ is less than $M$.

*Theorem 2: Sparseness of the inhomogeneous solution*
Consider the set of $N$ $M$-dimensional vectors $\vec{\Omega}_i = \left(\vec{W}_i, \theta_i\right)$. Assume that these vectors are of general position, i.e. any subset of $M+1$ of these vectors is linearly independent. Then in the minimum of the inhomogeneous Lyapunov function (15) no more than $M$ granule cells can be simultaneously active.
*Proof:* For any active granule cell the following equation is valid:

$$\partial \mathcal{L}/\partial a_i = -\sum_m W_{mi} \Delta x_m + \theta_i = 0. \qquad (16)$$

Assume that more than $M$ granule cells are active. Then we have at least $M+1$ such equations for $M$ unknowns $\Delta x_m$. Such a system in general case (if $M+1$ corresponding vectors $\vec{\Omega}_i$ are independent) is inconsistent and has no solution. Thus the number of coactive granule cells cannot exceed $M$.

Note: Consider the case of small but non-zero thresholds of firing of granule cells $\theta$. In this case two regimes can be distinguished. If vector $\vec{x}$ can be expanded in terms of vectors $\vec{W}_i$ with positive coefficients, the firing rates of $M$ granule cells are generally $\sim 1$ but the responses of mitral cells are small ($\sim \theta$). This is the regime of sparse overcomplete representation. If the glomerular input vector $\vec{x}$ cannot be represented as a superposition of granule cell weights $\vec{W}_i$ with positive coefficients (incomplete representation), the responses of cells are essentially (ignoring contributions $\sim \theta$) given by the solution of homogeneous problem (14), which explains our attention to this problem. In this case, according to Theorem 1, fewer than $M$ GCs have large firing rates and only one has a small firing rate ($\sim \theta$).

**Coactive granule cells in the case of sparse connectivity.** Assume that the weight matrix between granule and mitral cells is sparse, which implies that out of $M$ matrix elements $W_{mi}$ for the $i$-th GC only $K \ll M$ are non-zero. This corresponds to the situation when the GC can contact a limited subset of only $K$ mitral cells. Here we will calculate the number of GCs that are coactive in the case of sparse connectivity. Clearly, fewer than $M$ GCs have to be coactive according to Theorem 2. We argue however that the number of coactive GC can be substantially smaller than $M$.

With $\theta_i = 0$ (16) leads to the following approximate equation that is only valid for non-zero $a_i$-s

$$K a_i + (K^2/M) \sum_j a_j \approx p_i, \qquad (17)$$

where $p_i = \vec{W}_i \cdot \vec{x}$ is the projection of the input vector on synaptic weight. We can find therefore the activities of the GCs approximately as functions of the threshold $T$: $a_i = (p_i - T)/K$, $T = SK^2/M$, where $S = \sum_i a_i$. We can replace summation in the last equation with integration over the distribution of projections, $\rho(p) = N \cdot \exp[-(p-K)^2/K]/\sqrt{\pi K}$ (we will assume that the sparseness of input vectors is 50% for simplicity). For $T - K \gg \sqrt{K}$ we obtain

$$S(T) \sim N\sqrt{K} \frac{\exp\left[-(T-K)^2/K\right]}{(T-K)^2} \quad (18)$$

This equation has to be solved self-consistently with $T = SK^2/M$ to obtain the threshold for the input projection $T$ for granule cells to be active. Because the number of coactive granule cells can also be found from integrating $\rho(p)$ we have

$$N_G(T) \sim S \cdot (T-K) \sim M\sqrt{\ln N/K}. \quad (19)$$

## REFERENCES

Adrian ED (1950) The electrical activity of the mammalian olfactory bulb. EEG Clin Neurophysiol 2:377-388.

Arevian AC, Kapoor V, Urban NN (2008) Activity-dependent gating of lateral inhibition in the mouse olfactory bulb. Nat Neurosci 11:80-87.

Baraniuk RG (2007) Compressive sensing. Ieee Signal Proc Mag 24:118-+.

Cury K, Uchida (2009) Odor coding based on spike timing with respect to respiration-coupled oscillations during active sampling. In: Society for Neuroscience annual meeting. Chicago.

Davis BJ, Macrides F (1981) The organization of centrifugal projections from the anterior olfactory nucleus, ventral hippocampal rudiment, and piriform cortex to the main olfactory bulb in the hamster: an autoradiographic study. J Comp Neurol 203:475-493.

Doucette W, Restrepo D (2008) Profound context-dependent plasticity of mitral cell responses in olfactory bulb. PLoS Biol 6:e258.

Field D (1994) What is the goal of sensory processing? Neural Computation 6.

Fuentes RA, Aguilar MI, Aylwin ML, Maldonado PE (2008) Neuronal activity of mitral-tufted cells in awake rats during passive and active odorant stimulation. J Neurophysiol 100:422-430.

Hertz J, Krogh A, Palmer RG (1991) Introduction to the theory of neural computation. Cambridge, Mass.: Perseus.

Hopfield JJ (1982) Neural networks and physical systems with emergent collective computational abilities. Proc Natl Acad Sci U S A 79:2554-2558.

Hopfield JJ (1984) Neurons with graded response have collective computational properties like those of two-state neurons. Proc Natl Acad Sci U S A 81:3088-3092.

Kay LM, Laurent G (1999) Odor- and context-dependent modulation of mitral cell activity in behaving rats. Nat Neurosci 2:1003-1009.

Lee DD, Seung HS (1997) Unsupervised learning by convex and conic coding. In: Advances in Neural Information Processing Systems 9 pp 515-521.

Lennie P (2003) The cost of cortical computation. Curr Biol 13:493-497.

Lledo PM, Gheusi G, Vincent JD (2005) Information processing in the mammalian olfactory system. Physiol Rev 85:281-317.

Luskin MB, Price JL (1983) The topographic organization of associational fibers of the olfactory system in the rat, including centrifugal fibers to the olfactory bulb. J Comp Neurol 216:264-291.

Moulton DG (1963) Electrical activity in the olfactory system of rabbits with indwelling electrodes. In: Olfaction and Taste I (Zotterman Y, ed), pp 71-84. Oxford: Pergamon Press.

Olshausen BA, Field DJ (2004) Sparse coding of sensory inputs. Current Opinion in Neurobiology 14:481-487.

Rinberg D, Koulakov A, Gelperin A (2006) Sparse odor coding in awake behaving mice. J Neurosci 26:8857-8865.

Rozell CJ, Johnson DH, Baraniuk RG, Olshausen BA (2008) Sparse coding via thresholding and local competition in neural circuits. Neural Comput 20:2526-2563.

Seung HS, Richardson TJ, Lagarias JC, Hopfield JJ (1998) Minimax and Hamiltonian dynamics of excitatory-inhibitory networks. In: Advances in neural information processing systems 10, pp 329 - 335.

Shepherd GM, Chen WR, Greer CA (2004) Olfactory Bulb. In: The synaptic organization of the brain (Shepherd GM, ed), pp 165-216. New York: Oxford University Press.

Shepherd GM, Chen WR, Willhite D, Migliore M, Greer CA (2007) The olfactory granule cell: from classical enigma to central role in olfactory processing. Brain Res Rev 55:373-382.

Shusterman R, Koulakov A, Rinberg D (2009) The temporal structure of mitral/tufted cell odor responses in the olfactory bulb of the awake mouse. In: Society for Neuroscience annual meeting. Chicago.

Tolley LK, Bedi KS (1994) Undernutrition during early life does not affect the number of granule cells in the rat olfactory bulb. J Comp Neurol 348:343-350.